\documentclass[twocolumn,showpacs,preprintnumbers,amsmath,amssymb, prl]{revtex4}

\usepackage{stmaryrd}
\usepackage{txfonts}
\usepackage{amssymb}
\usepackage{mathrsfs}
\usepackage{graphicx}% Include figure files
\usepackage{dcolumn}% Align table columns on decimal point
\usepackage{bm}% bold math
\usepackage{epsfig}

\begin{document}

\title{Response and Amplification of Terahertz Electromagnetic Waves in Intrinsic Josephson Junctions of Layered High-$T_c$ Superconductor}

\author{Shi-Zeng Lin and Xiao Hu}

\affiliation{\(^{1}\)WPI Center for Materials
Nanoarchitectonics, National Institute for Materials Science, Tsukuba 305-0044, Japan\\
\(^{2}\)Japan Science and Technology Agency, 4-1-8 Honcho,
Kawaguchi, Saitama 332-0012, Japan}

\date{\today}

\begin{abstract}
We investigate the response of a stack of intrinsic Josephson
junctions (IJJs) to terahertz (THz) electromagnetic (EM)
irradiation. A significant amplification of the EM wave can be
achieved by the IJJs stack when the incident frequency equals to
one of the cavity frequencies. The irradiation excites $\pi$ phase kinks
in the junctions, which stimulate the cavity resonance when the bias
voltage is tuned. A large amount of dc energy is then pumped into
the Josephson plasma oscillation, and the incident wave gets
amplified. From the profound current step in \emph{IV}
characteristics induced at the cavity resonance, the system can also
be used for detection of the THz wave.

\end{abstract}

\pacs{74.50.+r, 74.25.Gz, 85.25.Cp}
 \maketitle

It has been known for a long time that Josephson junctions can be used as
oscillator, amplifier and detector for electromagnetic (EM) wave\cite{BaroneBook}.
The operating frequency of these devices made of conventional low-temperature
superconductors is below terahertz (THz) due to the small superconducting
energy gap. The discovery of intrinsic Josephson effect in layered high-$T_c$
superconductors\cite{Kleiner92}, such as $\rm{Bi_2Sr_2CaCu_2O_{8+\delta}}$(BSCCO),
has extended the frequency to the THz band, where EM waves have
potential for wide applications\cite{Ferguson02,Tonouchi07}, and thus has stimulated
intensive research activities in the field \cite{HuReview09}.

A breakthrough in generating coherent THz emission has been achieved
recently based on a mesa structure of BSCCO single crystal
\cite{Ozyuzer07}. Due to the thickness much smaller than the wave
length of EM wave, the IJJs stack itself forms a cavity,
synchronizes the plasma oscillation and radiates coherent THz wave
at the cavity resonance \cite{Bulaevskii06PRL}. The dynamics of the
superconductivity phase has been addressed theoretically
\cite{szlin08b,Koshelev08b} that $\pm\pi$ phase kinks are developed in the
junctions, which couple the dc bias to the standing wave and make
the cavity resonance possible.

In applications THz detector and amplifier are as important as
generator. Shapiro steps \cite{Shapiro63} were observed
in IJJs stacks of small BSCCO mesas under
THz irradiation \cite{Doh00,Wang01,Latyshev01,Bae08}, which can be
used for detection. To the best of our knowledge, no experiment on amplification of THz waves based
on IJJs has been reported so far. Now with the
success of generation of coherent THz wave at cavity resonance
\cite{Ozyuzer07}, it is intriguing to explore the possibility of the
same setup for the usage of amplification and detection of THz
waves, with the expectation that the system exhibiting a cavity
resonance in the THz band responds more sensitively to an incident wave
than short junctions reported in literatures.

When a stack of IJJs is irradiated by an EM wave, the transmitted wave excites
Josephson plasma oscillations inside the IJJs. The incident wave can be
either damped or amplified according to the detailed compensation
between dissipations caused by quasiparticles and the power supply
from the bias voltage, which, in turn, is governed by the phase
dynamics in the stack of IJJs.

By investigating the inductively coupled sine-Gordon equations
under appropriate boundary condition taking into account the THz EM
irradiation, we show in the present article that with the IJJs stack
one can achieve a significant amplification of the input
wave with frequency equal to the one of the cavity frequencies. Tuning the bias
voltage, $\pi$ phase kinks are created in the junctions, which pumps
a large amount of dc energy into the Josephson plasma oscillation
due to the cavity resonance. The profound current step in \emph{IV}
characteristics induced at the cavity resonance signals the
existence incident THz wave, and thus can be used for detection.

\begin{figure}[t]
\psfig{figure=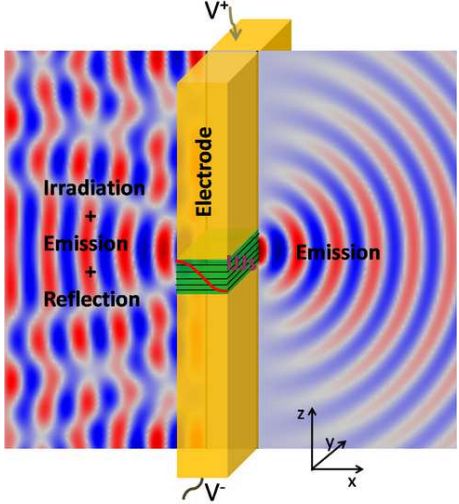,width=6cm} \caption{\label{f1}(Color
online) A stack of IJJs sandwiched by two electrodes and subjected
to irradiation. The whole setup is immersed in a dielectric material
with dielectric constant $\epsilon _d$.}
\end{figure}

The setup is shown in
Fig.~\ref{f1}, where a stack of IJJs are sandwiched by two ideal
conductors with infinite thickness. These two conductors prevent the
interference between EM waves from the two edges of IJJs stack. The
left side of IJJs is exposed to irradiation. We assume that the IJJs
are infinitely long in the $y$ direction, and thus the problem reduces
to two dimensions with sizes $L_x$ and $L_z$. This setup is
similar to the one proposed in Ref.\cite{Bulaevskii08}, except for
the lateral size $L_x\simeq 100 \mu$m which contains cavity modes
in the THz regime.

The dynamics of the gauge invariant phase difference in IJJs is
described by the inductively coupled sine-Gordon
equations\cite{Sakai93,Bulaevskii96,szlin08b,HuReview09}
\begin{equation}\label{eq1}
\partial _x^2 P_l  = (1 - \zeta \Delta_{\rm{d}})[\sin P_l  +
\beta \partial _t P_l  + \partial _t^2 P_l  - J_{{\rm{ext}}}
],
\end{equation}
where $P_l$ is the gauge invariant phase difference at the $l$-th
junction, $\beta\equiv4\pi\sigma_c\lambda_c/c\sqrt{\varepsilon_c}$
the normalized $c$-axis conductivity, $\zeta =(\lambda
_{ab}/s)^2$ the inductive coupling; ${{\varepsilon
}_{c}}$ is the dielectric constant and ${{\sigma }_{c}}$ is the
conductivity along the $c$-axis, and $s$ is the lattice period in
the $c$-direction; $c$ is the light velocity in vacuum; ${{\lambda
}_{c}}$ and ${{\lambda }_{ab}}$ are the penetration depths along the
$ab$-axis and $c$-axis respectively. In Eq.(\ref{eq1}), the lateral
space is normalized by ${{\lambda }_{c}}$, time by the Josephson
plasma frequency
${{\omega }_{J}}={c}/{{{\lambda }_{c}}\sqrt{{{\varepsilon }_{c}}}}$,
and the external current $J_{\rm{ext}}$ by the Josephson critical
current ${{J}_{\rm{c}}}$ \cite{szlin08}. $\Delta_{\rm{d}}$ is the
second-order difference operator defined as $\Delta_{\rm{d}} f_l\equiv
f_{l+1}+f_{l-1}-2f_l$. We adopt $\beta=0.02$ and $\zeta=7.1\times
10^4$, which are typical for BSCCO \cite{szlin09a}. The physics discussed below is valid in a stack of IJJs with huge $\zeta$, which has not yet been achieved in artificial Josephson junction stacks\cite{Ustinov93,Sakai93}.

In the absence of irradiation, the so-called dynamic boundary
condition (DBC) was derived based on Maxwell equations
\cite{Bulaevskii06PRL,Koshelev08}. It is easy to generalize the DBC
to incorporate irradiation because Maxwell equations are linear.
Assuming the incident wave is a plane wave with the electric
field polarized along the $z$ axis and the propagation direction
normal to the left edge of IJJs, the total electric field on the
left side is
\begin{equation}\label{eq1a}
\begin{array}{l}
 E_z(x, z, t)=E_z^{\rm{i}} (\omega)\exp [i( - \sqrt{\epsilon _d}\omega x + \omega t+\theta)]\\
 +\int dk_x E_z^{\rm{o}} (\omega ,k_x ,k_z )\exp [i( - k_x x + k_z z + \omega t)],\\
 \end{array}
\end{equation}
where $k^2_x+k^2_z=\omega^2$ with $E_z^{\rm{o}}$ the outgoing wave
comprising the emitted and reflected waves, $E_z^{\rm{i}}$ the
incident wave with a relative phase difference $\theta$ to the
Josephson plasma oscillation inside IJJs, $\omega$ the frequency,
and $\epsilon _d$ the normalized dielectric constant of the
dielectric medium coupled to the IJJs \cite{szlin08}. For simplicity of analysis, we concentrate in Eq.(\ref{eq1a})
on the case that the frequency of Josephson plasma determined by
the bias voltage according to the ac Josephson relation is equal
to the incident frequency, since otherwise the response of IJJs is
very small.

The EM wave at the right edge comes only from emission. The generalized
boundary conditions for the oscillating electromagnetic fields in
the real space and frequency domain are given by\cite{Bulaevskii08}
\begin{equation}\label{eq2}
B_y (x=0,\omega ) = \frac{E_z(0,\omega)}{Z(\omega)}-2\sqrt {\epsilon _d }E_z^{\rm{i}}(0, \omega)\exp(i\theta),
\end{equation}
\begin{equation}\label{eq3}
B_y (x=L_x,\omega ) = -\frac{E_z(L_x,\omega)}{Z(\omega )},
\end{equation}
where $B_y$ and $E_z$ are the total electric and magnetic fields,
and $Z(\omega )=2/\left\{ {k_\omega L_z \sqrt {\epsilon _d }\left[
{1 + \frac{{2i}}{\pi }\ln \frac{5.03}{{k_\omega L_z }}} \right]}
\right\}$ with $k_{\omega}\equiv\omega\sqrt{\epsilon _d}$ is the
impedance \cite{Bulaevskii06PRL}. The power of the incident wave is $S_{\rm i}=\sqrt{\epsilon _d}(E_z^{\rm i})^2/2$. Since the thickness of the IJJs stack used in experiments
is $L_z=1 \mu$m, much smaller than the wave length in the THz band, the
electromagnetic fields are uniform in the $z$ direction in the IJJs for THz
waves as in Eqs.~(\ref{eq2}) and (\ref{eq3}). There exists a
significant impedance mismatch between the IJJs and the dielectric
medium, which is crucial for the cavity resonances. With the
relations $(1-\zeta\Delta_{\rm{d}})B_l^y=\partial_x P_l$ and
$E_l^z=\partial_t P_l$,\cite{szlin09a} one obtains the boundary condition for the
oscillating part of $P_l$.

A solution to Eq.(\ref{eq1}) is given intuitively by
\begin{equation}\label{eq4}
P_l(x,t) = \omega t + {\mathop{\rm Re}\nolimits} [ - ig(x)\exp (i\omega t)],
\end{equation}
where the first term at the right-hand side (r.h.s) is the uniform
rotating phase according to the ac Josephson relation and the second
term is the plasma oscillation. The spatial modulation of plasma
oscillation is induced by both radiation and irradiation. We
consider the region of small plasma oscillation $|g(x)|<1$. From terms
with time dependence $\exp(\pm i\omega t)$, we obtain the equation for
$g(x)$ by substituting Eq.(\ref{eq4}) into Eq. (\ref{eq1})
\begin{equation}\label{eq5}
\partial _x^2 g(x)=1+i\beta \omega g(x)-g(x)\omega ^2.
\end{equation}
Equation (\ref{eq5}) has the solution
\begin{equation}\label{gx}
g(x) = A + a\exp (iqx) + b\exp ( - iqx)
\end{equation}
with $A = 1/(\omega ^2 - i\beta \omega )$ and $q\approx\omega$ for
weak damping $\beta\ll 1$ as in the case of BSCCO system.
The first term $A$ in Eq.~(\ref{gx}) represents the uniform plasma
oscillation, and the other two terms are propagating waves due to
radiation and irradiation, with the two coefficients $a$ and $b$
determined by the boundary condition Eqs.~(\ref{eq2}) and (\ref{eq3}).
\begin{figure}[t]
\psfig{figure=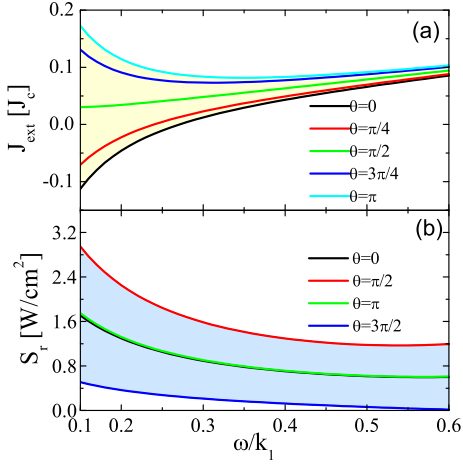,width=7cm}
\caption{\label{f2}(Color online). (a) \emph{IV} characteristics and (b) radiation power
at the right edge of the state described by Eq. (\ref{eq4}) for the incident wave
of $S_i=141\rm{W/cm^2}$ at several typical phases $\theta$. The light yellow regime
in between the maximal and minimal current associated with $\theta=0$
and $\theta=\pi$ is the width of the first Shapiro step. The results are
obtained with $L_x = 80\rm{\mu m}$, $L_z= 1\rm{\mu m}$ and $\epsilon_d=0.1$ similar to those in
the experiments \cite{Ozyuzer07}.}
\end{figure}

The \emph{IV} characteristics is derived from the
current conservation relation
\begin{equation}\label{eqkink4}
J_{\rm{ext}}  = \beta \omega  + \left\langle{ \sin P_l }\right\rangle _{xt}
\end{equation}
where $\left\langle \cdots \right\rangle_{xt}$ denotes the average
over space and time. Besides the normal current due to the
quasiparticles $J_{\rm{n}}=\beta\omega$, the total dc current
has two contributions from the plasma oscillation due to
the nonlinearity of the dc Josephson effect:
$J_{\rm{p}}=\beta/[2(\omega^3 + \beta ^2\omega )]$ and
$J_{\rm{w}}= \{a[1 - \exp (i\omega L_x)]-b[1 - \exp ( - i\omega
L_x)]\}/(2\omega L_x)$
associated with the uniform and nonuniform parts of plasma oscillation
in Eq.(\ref{gx}). The \emph{IV} curves for different
$\theta$'s are displayed in Fig.~\ref{f2}(a) for the incident
wave $S_{\rm i}=141\rm{W/cm^2}$. When one fixes the
voltage satisfying the phase-locking relation and sweeps the current,
the relative phase $\theta$ adjusts itself to match the current, which
traces out the Shapiro steps \cite{Shapiro63}.
Zero-crossing Shapiro steps \cite{Kautz96} occur at small voltages.
For $1/Z\ll \omega L_x\ll 1$, $J_{\rm{w}}$ is given explicitly as
\begin{equation}\label{eq6}
J_{\rm{w}}=\text{Re} \left[\frac{ 1}{L_x \omega^3 Z}
-\frac{E_z^{\rm i} e^{i \theta} \sqrt{\epsilon _d}}{L_x \omega ^2}\right].
\end{equation}
$J_{\rm{w}}$ is maximized (minimized) at $\theta=\pi$ ($\theta=0$), and the height of the Shapiro step is given
by $J_{\rm{s}}={2 E_z^{\rm i} \sqrt{\epsilon _d}}/{L_x \omega ^2}$. In the region where
$\Delta J_{\rm{w}}\equiv J_{\rm{w}}(E_z^{\rm i}>0)-J_{\rm{w}}(E_z^{\rm i}=0)>0$,
a dc power $\Delta J_{\rm{w}}\omega$ is converted into emission. When
$J_{\rm{w}}<0$, the incident EM wave is converted into dc power
and charges the IJJs, and the IJJs effectively work as a battery.

The radiation powers measured by the Poynting vector
satisfy the power balance condition
$S_r-S_l+P_{d}=J_{\rm{ext}}\omega$ with $P_d$ the dissipation
caused by the quasiparticles, and $S_l$ and
$S_r$ the radiation at the left and right edge respectively. The emission
at the right edge is depicted in Fig.~\ref{f2}(b) for $S_{\rm i}=141\rm{W/cm^2}$.
It is clear that the emission is always very weak because of lack of an
efficient way to pump energy into plasma oscillation in this state.
In the same limit, the radiation power is given by
\begin{equation}\label{eq7}
S_r=\text{Re}\left[\frac{\omega ^2}{2Z^*}\right]\left|A-
\frac{2 i E_z^{\rm i} e^{i \theta}\sqrt{\epsilon _d}}{L_x \omega ^2}\right|^2.
\end{equation}
It is clear that the emission comprises of spontaneous one
$S_{\rm{sp}}\sim |A|^2$, the one caused by transmitted
wave $S_{\rm{tw}}\sim |E_z^i|^2$ and the stimulated one
$S_{\rm{st}}\sim |E_z^i A|$.

\begin{figure}[t]
\psfig{figure=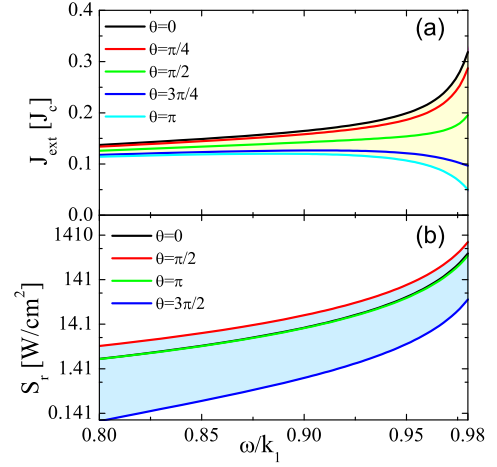,width=7cm}
\caption{\label{f3}(Color online). (a) \emph{IV} characteristics and (b) radiation power when the $\pi$ phase kinks are excited near the cavity resonance. Other parameters are the same as those in Fig. \ref{f2}.}
\end{figure}

The state given in Eq.(\ref{eq4}) is found to become unstable when
the incident frequency and the bias voltage get close to
$\omega=k_1\equiv\pi/L_x$,
where the cavity mode $g(x)\approx A_1\cos(k_1x)$ with $A_1>1$ is
induced by the irradiation (without losing generality, here we
consider the first cavity mode). Instead of Eq.(\ref{eq4}) the phase
dynamics should then be described by \cite{szlin08b}
\begin{equation}\label{eqkink1}
P_l (x,t) = \omega t + P_l^s (x) + {\mathop{\rm Re}\nolimits} [ - ig(x)\exp (i\omega t)],
\end{equation}
where $P_l^s$ is the static phase kink associated with the
cavity mode. Substituting Eq.~(\ref{eqkink1}) into Eq.~(\ref{eq1}),
we obtain the following equations
\begin{equation}\label{eqkink2}
\partial _x^2 g(x)=(1 - \zeta \Delta_{\rm d} )\exp (iP_l^s ) + i\beta \omega g(x)-g(x)\omega ^2
\end{equation}
and
\begin{equation}\label{eqkink3}
\partial _x^2 P_l^s  = \frac{i}{2}\zeta \Delta_{\rm{d}} g(x)\exp ( - iP_l^s ).
\end{equation}
Since the Josephson plasma should take the form
\begin{equation}\label{gxkink}
g(x) = A_1\cos k_1 x + a\exp (iqx) + b\exp ( - iqx),
\end{equation}
where $q\thickapprox\omega$, we arrive at
 $A_1=F_1/(ik_1^2 -i\omega^2 -\beta\omega)$
with $F_1 = \frac{{ - 2i}}{L_x}\int\limits_0^{L_x}
{(1 - \zeta \Delta_{\rm d} )\exp (iP_l^{s} )\cos (k_1 x)dx}$.
Due to that the dominant term in Eq.(\ref{gxkink})
$g(x)\approx A_1\cos(k_1x)$ is antisymmetric with respect to $x=L_x/2$,
Eq. (\ref{eqkink3}) has
a solution of $\pi$ kinks alternatingly piled up in the $c$-axis\cite{szlin08b}. Because of the huge inductive coupling
$\zeta\approx 10^5$ in BSCCO, the
phase kinks render themselves as step functions, and are stable against
the radiation and irradiation provided $|a|, |b|<|A_1|$.
The width of a kink should be smaller than the junction width
$1/\sqrt{|A_1|\zeta}< L_x$, which gives an estimate on the regime
where the kink state is stable.

\begin{figure}[t]
\psfig{figure=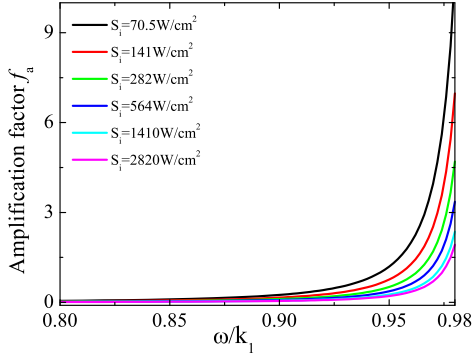,width=7cm} \caption{\label{f4}(Color online). Amplification
factor for several typical incident powers near the cavity resonance.}
\end{figure}

Although it has been revealed theoretically that the
$\pi$ kink state is ideal for generating strong terahertz electromagnetic
waves \cite{szlin08b,HuReview09}, the dynamic process to realize the state was not clear.
The present study indicates that irradiating the junction stack by
an incident wave can stimulate the $\pi$ kink state.

The dc supercurrent induced by the plasma oscillation in the kink state
can be evaluated by
$\left\langle {{ - ig(x)}\exp ( - iP_l^s )/2} \right\rangle _x$.
It includes the current associated with the plasma oscillations at the cavity mode
$J_{\rm{p}}=4\beta \omega /\{\pi^2[(k_1^2 -\omega ^2 )^2
+\beta ^2\omega ^2]\}$, and that with radiation and irradiation
$J_{\rm{w}}= (a\left[\exp(i\omega
L_x/2)-1\right]^2-b\left[\exp(-i\omega L_x/2)-1\right]^2)/(2\omega L_x)$.
The \emph{IV} characteristics with the irradiation of $S_{\rm i}=141\rm{W/cm^2}$
is given in Fig.~\ref{f3}(a). For  $1/Z\ll k_1-\omega\ll1$ and $(k_1^2-\omega^2)\gg\beta\omega$, we have
\begin{equation}\label{eqkink7}
J_{\rm{w}}  = \text{Re}\left[\frac{8}{L_x \pi^2 Z(k_1^2-\omega^2)(k_1-\omega) }+\frac{2 E_z^{\rm{i}} e^{i\theta }\sqrt{\epsilon _d}}{ \pi ^2(k_1-\omega)}\right].
\end{equation}
Because of the $\pi$ phase kinks, now $J_{\rm{w}}$ is maximized (minimized) at $\theta=0$ ($\theta=\pi$) in contrast to Eq. (\ref{eq6}). The height of the Shapiro step is
$J_{\rm{s}}=4 E_z^{\rm i}\sqrt{\epsilon _d}/[\pi ^2(k_1-\omega)]$.
As is well known, Shapiro steps are suppressed by internal modes in
single junctions. It is the same case for a IJJs stack if the state
is uniform along the $c$-axis, since Eqs.(\ref{eq1}) are decoupled.
Therefore, the appearance of a Shapiro step at the
cavity resonance can be used as an exclusive detection of the
$\pi$ kink state in a stack of IJJs.

The radiation power at the right edge is depicted in Fig.~\ref{f3}(b)
for $S_{\rm i}=141\rm{W/cm^2}$. It is clear that the input wave is enhanced
significantly near the cavity frequency, where the $\pi$ kinks stimulated
by the irradiation pump a large amount of dc power from the dc bias into
Josephson plasma oscillation.

The radiation power at the right edge in the same limit
is given by
\begin{equation}\label{eqkink5}
S_r =\text{Re}\left[\frac{\omega ^2}{2Z^*}\right]
\left|-\frac{4}{\pi(k_1^2-\omega^2)}+\frac{2i E_z^{\rm i} e^{i \theta}\sqrt{\epsilon _d}}{\pi(k_1-\omega)
}\right|^2.
\end{equation}
An amplification factor can be defined by the maximal value
of the ratio $S_r/S_{\rm i}$ with respect to the phase $\theta$ for a
given $S_{\rm i}$. For the $\pi$ kink state, one has
 \begin{equation}\label{eqkink6}
f_{\rm{a}} =\text{Re}\left[\frac{\omega ^2}{\sqrt{\epsilon _d}Z^*}\right]\left|\frac{4}{\pi(k_1^2-\omega^2)E_z^{\rm i}}+\frac{2\sqrt{\epsilon _d}}{\pi(k_1-\omega)
}\right|^2,
\end{equation}
which is achieved at $\theta=\pi/2$. As displayed in Fig.~\ref{f4}, the amplification factor reaches
its maximum at $\omega=k_1$. An incident wave
of power of $141\rm{W/cm^2}$ can be amplified by one order of
magnitude. The amplification factor decreases with the power of
incident wave, and the maximum power which can be amplified by this
technique is estimated as $3000\rm{W/cm^2}$.

In a single junction, the presence of irradiation may cause chaotic dynamics in a certain parameter space\cite{Kautz96}, which is harmful for applications. The chaos can be avoided when the frequency of the incident wave is much larger than the Josephson plasma frequency\cite{Kautz96,Wang01}, which is fulfilled in a IJJs stack of length smaller than $\lambda_c$.

In conclusion, simultaneously shining a terahertz electromagnetic wave and biasing
a dc voltage on a stack of intrinsic Josephson junctions stimulates
the standing wave of Josephson plasma, which develops $\pi$ phase
kinks in the junctions, when the frequency equals to one of the
cavity frequencies of the junction stack. At the cavity resonance,
the rotating $\pi$ kinks pump a large amount of dc energy into
Josephson plasma oscillation, and the incident wave gets amplified.
The maximal radiation power reached by this terahertz amplifier is
estimated as $3000{\rm W/cm^2}$. Since the strong plasma oscillation
induces a large dc supercurrent at the cavity resonance, the system
can work as a terahertz detector. The response of the system to irradiation depends on the spatial structure of the superconductivity phase, thus the phase dynamics may be probed by the irradiation.

This work was supported by WPI Initiative on Materials Nanoarchitronics,
MEXT, Japan and CREST-JST Japan. Calculations were performed on the Numerical Materials Simulator (SGI Altix supercomputer) in NIMS.

%\bibliography{reference}

\end{document}